 \documentclass[pdflatex,sn-nature]{sn-jnl}

\usepackage{pdfpages}
\usepackage{graphicx}%
\usepackage{multirow}%
\usepackage{amsmath,amssymb,amsfonts}%
\usepackage{amsthm}%
\usepackage{mathrsfs}%
\usepackage[title]{appendix}%
\usepackage{xcolor}%
\usepackage{textcomp}%
\usepackage{manyfoot}%
\usepackage{booktabs}%
\usepackage{algorithm}%
\usepackage{algorithmicx}%
\usepackage{algpseudocode}%
\usepackage{listings}%

\usepackage{amsmath}
\usepackage{amsfonts}
\usepackage{amssymb}

\usepackage{revsymb}
\usepackage{graphicx}
\usepackage{siunitx}
\usepackage{bm}
\usepackage{dcolumn}
\usepackage{multirow}
\usepackage{graphicx}
\graphicspath{{figures/}}
\usepackage{epsfig}
\usepackage{float}
\usepackage{tabularx}
\usepackage{booktabs}
\usepackage{xcolor}

\def\ket#1{|\,#1 \,\rangle}

\def\eref#1{Eq.~(\ref{#1})}
\def\tref#1{Table~\ref{#1}}

\theoremstyle{thmstyleone}%
\theoremstyle{thmstyletwo}%

\theoremstyle{thmstylethree}%

\begin{document}

\title{Lu$^+$ optical frequency references with accuracy verified at the $19^\mathrm{th}$ digit}


\author*[1,2]{\fnm{K.J.} \sur{Arnold}}\email{cqtkja@nus.edu.sg}
\equalcont{These authors contributed equally to this work.}

\author[1]{\fnm{M. D. K.} \sur{Lee}}
\equalcont{These authors contributed equally to this work.}

\author[1]{\fnm{Qi} \sur{Zhao}}

\author[1]{\fnm{Qin} \sur{Qichen}}

\author[1]{\fnm{Zhao} \sur{Zhang}}

\author[1]{\fnm{N.} \sur{Jayjong}}

\author*[1,3]{\fnm{M. D.} \sur{Barrett}}\email{phybmd@nus.edu.sg}

\affil[1]{\orgdiv{Centre for Quantum Technologies}, \orgname{National University of Singapore}, \orgaddress{\street{3 Science Drive 2}, \city{Singapore}, \postcode{117543}, \country{Singapore}}}

\affil[2]{\orgdiv{Temasek Laboratories}, \orgname{National University of Singapore}, \orgaddress{\street{5A Engineering Drive 1}, \city{Singapore}, \postcode{117411}, \country{Singapore}}}

\affil[3]{\orgdiv{Department of Physics}, \orgname{National University of Singapore}, \orgaddress{\street{2 Science Drive 3}, \city{Singapore}, \postcode{117551}, \country{Singapore}}}



\maketitle
\newpage

{\bf Optical atomic frequency references~\cite{ludlow2015optical} have far surpassed their microwave frequency predecessors, leading to an anticipated redefinition of the SI second~\cite{dimarcq2024roadmap}.  However, as the scientific community seeks consensus on a new standard, and several state-of-the-art optical standards now report evaluated fractional uncertainties below $10^{-18}$, verification by same-species comparisons to comparable levels remains an outstanding challenge. Here we report two $^\mathbf{176}$Lu$^+$ single-ion optical frequency references, each with evaluated fractional frequency uncertainty near $1\times10^{-19}$, which are directly compared via correlation spectroscopy and demonstrate agreement, with a measured relative frequency difference of $[-0.1\pm(5.7)_\mathrm{stat}\pm(1.0)_\mathrm{sys}]\times10^{-19}$, where `stat' and `sys' indicate the statistical and systematic uncertainty, respectively. Our optical reference has been comprehensively assessed with an evaluated uncertainty below $10^{-18}$ supported by a same-species comparison of independent systems to the $5.7\times10^{-19}$ level.  This accuracy, achieved in practical room-temperature systems, will contribute to improving international timekeeping and towards chronometric leveling~\cite{lion2017determination,mehlstaubler2018atomic} at the millimeter scale, as well as to tests of fundamental physics~\cite{safronova2018search} such as Lorentz invariance~\cite{sanner2019optical}, general relativity~\cite{takamoto2020test}, searches for dark matter~\cite{kennedy2020precision,filzinger2023improved}, and variation of fundamental constants~\cite{godun2014frequency,huntemann2014improved}. }

As the international scientific community seeks consensus on a new standard for redefinition of the SI second~\cite{dimarcq2024roadmap}, it is essential that the uncertainty budgets of optical standards are both rigorously evaluated and experimentally tested through comparison. While several state-of-the-art optical standards now report evaluated fractional uncertainties below $10^{-18}$~\cite{marshall2025high,aeppli2024clock,zhang2026liquid}, same-species comparisons to experimentally test these claims are relatively few~\cite{mcgrew2018atomic,sanner2019optical,takamoto2020test,zhiqiang2023}. Meanwhile inter-species comparisons at the highest precision have revealed significant inconsistencies relative to reported uncertainty budgets~\cite{collaboration2021frequency,aeppli2025atomic,dorscher2021optical,clock2024international}.   An evaluated uncertainty budget is a quantitatively testable prediction that two frequency standards of the same species, corrected for their evaluated systematic shifts, should agree to within their combined uncertainties. In this work we test that prediction directly by adhering to three criteria for a rigorous assessment of our optical frequency standard~\cite{arnold2024validating}: (i) a clear set of experimental measurements of atomic properties and environmental factors determining the systematic uncertainty budget; (ii) a same-species comparison to demonstrate validity of the uncertainty budget to the degree the measurement precision allows; and (iii) a stress test on any systematic that is substantially larger than the measurement precision of the comparison, i.e., a frequency comparison in which the systematic is deliberately shifted and observed to give the predicted shift in the clock frequency. Applying these criteria, we report the first comprehensive characterization of all known sources of systematic uncertainty for two independent $^{176}$Lu$^+$ single-ion optical references, each with evaluated systematic uncertainty near $1\times10^{-19}$, a fourfold improvement on the lowest previously reported~\cite{zhang2026liquid,marshall2025high}. Their agreement is verified at the level of $5.7\times10^{-19}$, limited by the comparison precision, after 200 hours of averaging. 
 
The exceptionally low systematic uncertainty is possible due to the relative insensitivity of the $^{176}$Lu$^+$ $^1S_0\leftrightarrow{}^3D_1$ transition to perturbations compared to other leading contenders. In particular, this transition has the lowest sensitivity to black-body radiation (BBR) and magnetic fields of any established clock system, and the large atomic mass makes it less susceptible to motional shifts than lighter atomic species. The complete uncertainty assessment, summarized in \tref{tab:sys}, builds on several recent advances, including an improved method for micromotion compensation~\cite{arnold2024enhanced}, evaluation of the quadrupole shift via $^{176}$Lu$^+$ microwave spectroscopy~\cite{lee2025quadrupole}, and evaluation of background gas collision effects~\cite{mbd_collisions}.

Frequency comparison at these extreme levels of precision on practical timescales requires exceptionally high stability, a figure of merit for which single trapped ion optical standards have historically lagged relative to neutral-atom optical lattice clocks~\cite{kim2025atomic,hinkley2013atomic,liu2025zero,oelker2019demonstration}. In this work, we employ correlation spectroscopy~\cite{chwalla2007precision,chou2011quantum,clements2020lifetime}, which rejects common-mode phase noise from a shared clock laser and enables interrogation times limited only by the atoms. The measured agreement therefore characterizes the atomic references themselves, independent of the performance of any particular clock laser. Here we demonstrate interrogation times up to ten seconds and a comparison measurement instability of $4.8\times10^{-16}(\tau/\mathrm{s})^{-1/2}$, nearly a factor of 3 improvement over previous work~\cite{zhiqiang2023} and within a factor of three of the best reported ratio measurement instabilities involving an ionic frequency standard realized by correlation spectroscopy~\cite{clements2020lifetime}, differential spectroscopy~\cite{kim2023improved}, or clock comparisons~\cite{aeppli2025atomic} with state-of-the-art laser stabilization~\cite{lee2026frequency}.

For $^{176}$Lu$^+$, a same-species comparison of independent systems provides exceptionally strong verification because all evaluated systematic effects --- except for the readily controlled quadratic Zeeman shift ---  are small compared to the statistical measurement precision. Consequently, even if these known effects were perfectly common-mode, they could not mask an error larger than the statistical uncertainty. To undermine the validity of the claimed performance, a hypothetical systematic shift would have to be both of unknown physical origin and common-mode across the two independent experimental setups.

\begin{figure*}[ht]
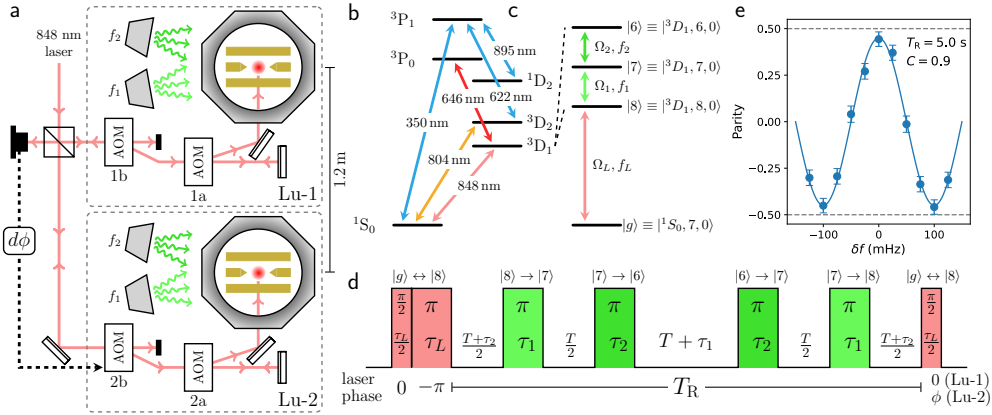

\centering
\caption{{\bf $^{176}$Lu$^+$ correlation spectroscopy.} (a) Simplified experimental scheme for correlation spectroscopy used to compare Lu-1 and Lu-2.  Acousto-optic modulators (AOMs) labeled 1a and 2a control the intensity, frequency, and relative phase of clock laser pulses delivered to the respective trapped ions, while AOM 2b actively compensates for differential path length fluctuations referenced to retroreflecting mirrors near the viewports of the respective vacuum chambers. Microwaves are delivered by horns external to the vacuum chambers. (b)  Atomic level structure of $^{176}$Lu$^+$ showing the wavelengths of transitions used. (c) The 848-nm optical clock transition and ${^3}D_{1}$ microwave clock transitions addressed in the clock interrogation sequence. $\Omega_\alpha$ and $f_\alpha$ denote the coupling strengths and frequencies for the fields driving the transitions indicated. (d) Clock interrogation sequence for hyperfine-averaged hyper-Ramsey spectroscopy. (e) Parity signal observed for a scan of the optical frequency difference with a Ramsey time $T_\mathrm{R} = 5\,$s.}
\label{fig:scheme}
\end{figure*}

\section*{Experimental System}
All optical clock systems use some form of averaging, in which the frequencies of two or more transitions are combined such that the average frequency is insensitive to perturbations from external fields.  Most commonly this is an average over one or more pairs of Zeeman lines, which cancels the linear Zeeman shift~\cite{bernard1998laser}. In the case of $^{176}$Lu$^+$, we employ averaging over the transitions from $\ket{g}\equiv \ket{^1S_0,7,0}$ to $\ket{F}\equiv \ket{^3D_1,F,0}$ for $F=6,7,$ and $8$ at the optical frequencies $f_F$. Conceptually, this hyperfine averaging technique~\cite{barrett2015developing} leads to an effective $J=0$ level for which the quadratic Zeeman and quadrupole shifts of the ($J=1$) $\ket{^3D_1,F,0}$ states are largely eliminated in the hyperfine average (HA) frequency, $f\equiv\frac{1}{3} (f_6+f_7+f_8)$. Rather than sequentially interrogate multiple optical transitions, the average can be conveniently realized by Ramsey spectroscopy on a single optical transition by using the hybrid microwave and optical pulse sequence~\cite{kaewuam2020hyperfine,zhiqiang2023}. As illustrated in Fig.~\ref{fig:scheme}({b and c}), the interrogation sequence consists of a Ramsey interrogation on the $\ket{g}$ to $\ket{8}$ transition and microwave transitions within the interrogation time to transfer population between the hyperfine states.  Timing is such that the effective time in each of the hyperfine states is the same.  Steering the laser to the central Ramsey fringe ensures the linear combination $f_L+\frac{1}{3}(2f_1+f_2)$ is the HA frequency $f$. This interrogation scheme has several advantages: exclusive use of first order field insensitive $m=0$ magnetic substates, the HA is realized every interrogation cycle, and it is readily extended to incorporate the hyper-Ramsey (HR)~\cite{yudin2010hyper,hyperRamsey2012peik} technique for suppression of the laser-induced ac-Stark shift by inclusion of an additional phase shifted optical $\pi$ pulse. The timing, pulse areas, and phases of the full hyperfine-average-hyper-Ramsey (HA-HR) sequence used in this work are illustrated in Fig.~\ref{fig:scheme}{d} with further details given in Methods.    

We compare single $^{176}$Lu$^+$ ions in two independent ion traps, denoted Lu-1 and Lu-2, via correlation spectroscopy~\cite{chwalla2007precision,chou2011quantum} which enables comparison of the references at high stability limited only by the differential atomic coherence rather than the laser coherence~\cite{clements2020lifetime}. The experimental setup is illustrated schematically in Fig.~\ref{fig:scheme}{a} and is similar to that described in previous work~\cite{zhiqiang2023}.  After initialization of both ions in the $\ket{g}$ state, the ions are simultaneously interrogated with the HA-HR sequence followed by detection of population in the $^3D_1$ state. For total Ramsey time $T_\mathrm{R}$ much longer than the laser coherence time, the expectation value of the parity, denoted $\Pi$, is given by~\cite{clements2020lifetime} 
\begin{equation}
\Pi(\phi) = \frac{C}{2} \cos(2\pi \delta fT_\mathrm{R}+\phi)
\end{equation}
where $\delta f  \equiv f^{(1)}-f^{(2)}$ is the difference in the HA-equivalent linear combination of laser and microwave field frequencies $f^{(k)} = f_L^{(k)} + \frac{1}{3}(2f_1^{(k)}+f_2^{(k)})$ used within the interrogation sequence for Lu-$k$, $\phi$ is a phase step applied only to the final optical $\frac{\pi}{2}$ pulse for Lu-2, and $C\equiv\max(\Pi)-\min(\Pi)\leq1$ is the parity fringe contrast. A typical frequency scan demonstrating minimal loss of parity fringe contrast for a Ramsey time $T_\mathrm{R} = 5\,$s is shown in Fig.~\ref{fig:scheme}{e}.

\begin{figure}[t]
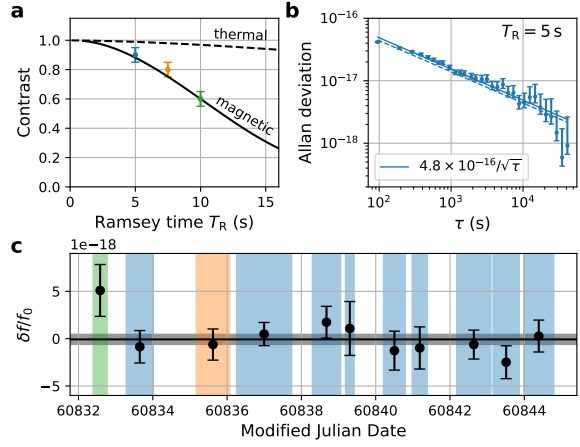

\centering
\caption{{\bf $^{176}$Lu$^+$ comparison results.} {\bf a,} Parity contrast inferred from the servo instability (points) compared to the simulated contrast loss due to ion heating (dashed black line) and due to uncorrelated $7\,$nT magnetic field flicker noise in both chambers (solid black line). {\bf b,} Allan deviation for the longest continuous measurement with $T_\mathrm{R}=5$ s. The dashed line is the projection noise limit for full parity contrast ($C=1$) and the solid line is observed instability $4.8\times10^{-16} (\tau/\mathrm{s})^{-1/2}$ corresponding to $C=0.9$. {\bf c,} Frequency difference (black points) from 11 measurements with a combined duration of 8.3 days over a 12.4 day interval (67\% uptime). Error bars represent the statistical uncertainties which are assumed to be projection noise limited. Shaded vertical bars show the measurement intervals, with the color indicating Ramsey time for $T_R=$  5 s (blue), 7.5 s (orange), and 10 s (green). The weighted mean frequency difference is $[-0.1\pm(5.7)_\mathrm{stat}\pm(1.0)_\mathrm{sys}]\times10^{-19}$ with reduced chi squared $\chi_\nu^2 = 0.75$ for 10 degrees of freedom. }
\label{fig:results}
\end{figure}

At the typical operating magnetic field of $B_0=0.1$ mT in both experiment chambers, we observe the contrast loss as a function of interrogation time shown in Fig.~\ref{fig:results}{a}. In contrast to previous work~\cite{zhiqiang2023}, we are no longer limited by thermal dephasing due to ion heating: the dashed line in Fig.~\ref{fig:results}{a} represents the simulated contrast loss due to thermal effects as evaluated for the current system. Although the HA frequency is extremely insensitive to magnetic fields, the individual component transitions are weakly susceptible to magnetic fields via the quadratic Zeeman shifts of the $\ket{^3D_1,F,m_F=0}$ states, for which the $\ket{6}$ and $\ket{8}$ states have linear magnetic sensitivities of $\approx \pm 4\, \mathrm{kHz/mT^2} B_0 \sim \pm 400\, \mathrm{Hz}/\mathrm{mT}$. Independent measurements on magnetic sensitive states indicate the ambient magnetic field has flicker instability of $7\,\mathrm{nT}$ on the timescales of interest for clock interrogation ($\sim 1-100\,\mathrm{s}$). The loss of contrast observed is consistent with simulations of uncorrelated magnetic field noise at this level for each ion, shown as the solid line in Fig.~\ref{fig:results}{a}. Coherence times on the microwave transitions of individual systems, their dependence on the dc magnetic field, and the observed contrast for HA-HR correlation spectroscopy are all consistent with uncorrelated magnetic field noise at this level.

\section*{Results}

The results of 11 comparison measurements totaling 200 hours are summarized in Fig.~\ref{fig:results}{c}.  With the exception of two initial measurements taken with Ramsey times of 7.5 s and 10 s, a Ramsey time of $T_R = 5$ s was used. In all cases the observed instability is consistent with the quantum projection noise limit after accounting for the fringe contrast. Fig.~\ref{fig:results}{b} shows the Allan deviation of the longest continuous run (37 hours, $T_R=5\,$s) demonstrating an instability of $4.8\times10^{-16}(\tau/\mathrm{s})^{-1/2}$. The weighted mean frequency difference of all 11 measurements is $[-0.1\pm(5.7)_\mathrm{stat}\pm(1.0)_\mathrm{sys}]\times10^{-19}$. 

\begin{table*}[t]
\caption{{\bf Clock uncertainty budget.} All values are in units $10^{-19}$ relative to the Lu$^+$ optical standard frequency. Some effects are common mode which is reflected in the lower uncertainty in the difference. Descriptions for each of these effects can be found in the main text and Methods. }
\label{tab:sys}
\begin{tabular*}{1.02\textwidth}{@{\extracolsep{\fill}} >{\quad}l r r r r r r <{\quad}}
\toprule[0.75pt]
 & \multicolumn{2}{c}{Lu-1} &  \multicolumn{2}{c}{Lu-2} & \multicolumn{2}{c}{Difference} \\
\cmidrule[0.5pt](lr){2-3} \cmidrule[0.5pt](lr){4-5} \cmidrule[0.5pt](lr){6-7}
Effect & \multicolumn{1}{c}{Shift} & \multicolumn{1}{c}{Unc.} & \multicolumn{1}{c}{Shift} & \multicolumn{1}{c}{Unc.} & \multicolumn{1}{c}{Shift} & \multicolumn{1}{r<{\quad}}{Unc.} \\  \hline
Quadratic Zeeman &-1383.69 &0.15 &-1383.69 &0.15 &0.00 &$<$0.01\\
AC Zeeman (rf) &0.27 &0.01 &0.52 &0.01 &-0.24 &0.02\\
AC Zeeman (microwave) &-0.03 &0.01 &-0.07 &0.01 &0.05 &0.01\\
Micromotion (excess) &-0.16 &0.13 &-0.14 &0.04 &-0.02 &0.13\\
Micromotion (phase) &0~~ &$<$0.01 &-2.42 &0.19 &2.42 &0.19\\
SODS (thermal)$^{\dagger}$ &-4.68 &0.31 &-5.04 &0.27 &0.37 &0.42\\
BBR &-13.9 &1.1~ &-13.8 &1.0~ &-0.18 &0.50\\
Microwave coupling &0~~ &0.08 &0~~ &0.52 &0~~ &0.52\\
Residual quadrupole &0.33 &0.03 &1.35 &0.14 &-1.02 &0.14\\
AC Stark (848 nm) &0~~ &0.04 &0~~ &0.25 &0~~ &0.25\\
AC Quadrupole (rf) &-0.02 &$<$0.01 &-0.01 &$<$0.01 &-0.01 &$<$0.01\\
Background collisions &0~~ &0.39 &0~~ &0.32 &0~~  &0.07\\
rf synthesis &-~~ &-~~ &-~~ &-~~ &0~~ &0.10\\
Differential phase chirp &-~~ &-~~ &-~~ &-~~ &-0.05 &$<$0.01\\
Gravitational &-~~ &-~~ &-~~ &-~~ &-4.31 &0.29\\\midrule
Total &-1401.9 &1.2~ &-1403.3 &1.3~ &-3.0~ &1.0~\\
\bottomrule[0.75pt]
$^{\dagger}$ For $T_R=5\,$s 
\end{tabular*}
\end{table*}

We evaluate a total systematic uncertainty of $1.2\times10^{-19}$ for Lu-1, $1.3\times10^{-19}$ for Lu-2, and $1.0\times10^{-19}$ for the difference, with individual contributions summarized in \tref{tab:sys}. The only systematic shifts with magnitude larger than $10^{-18}$ are the quadratic Zeeman shift and the BBR shift, which is a testament to the insensitivity of $^{176}$Lu$^+$ to perturbation.  We proceed with a brief discussion on the evaluation of systematic effects leaving the technical details to the Methods and Extended Data. 

The quadratic Zeeman shift is, by a factor of 100, the largest systematic and the only systematic more than an order of magnitude larger than our measurement precision.  It is characterized by the quadratic Zeeman coefficient for which a value of $\alpha_z =-4.892\,64(88)\,\mathrm{Hz}/\mathrm{mT}^{2}$ has been previously reported~\cite{zhiqiang2023}. Measurement of the quadratic Zeeman coefficient constitutes a stress test of the reference in accordance with our third criterion, as it calibrates changes in the reference as a function of applied magnetic field. In the interest of scientific rigor and since it is the largest systematic shift, $\alpha_z$ was remeasured in both the years 2024 and 2025, with both measurements being subsequent to a major lab renovation and replacement of the Lu-1 trap.  The later measurement data is shown in Extended Data Fig.~\ref{fig:qz}{a-b}.  There is no statistically significant deviation from a quadratic dependence, which highlights the effectiveness of hyperfine averaging operating over a wide range of magnetic fields. The three measurements of $\alpha_z$ are shown in Extended Data Fig.~\ref{fig:qz}{c}. We take the weighted mean value of $\alpha_z = -4.893\,27(51)\,\mathrm{Hz}/\mathrm{mT}^{2}$ where the statistical uncertainty has been scaled by the square root of the reduced chi-square statistic $\chi^2_\nu=1.85$. It is noted the probability $P[\chi^2_\nu>1.85]=0.16$ for two degrees of freedom and so the three measurements of $\alpha_z$ are statistically consistent.  For the high accuracy comparison at 0.1 mT,  uncertainty in $\alpha_z$ contributes $1.4\times10^{-20}$ uncertainty to the individual optical references, but this contribution is correlated in the comparison. The uncertainty in the difference of quadratic Zeeman shifts is determined from the magnetic field stability and is $<10^{-21}$ owing to the tracking and active steering of the magnetic fields in both chambers.

The blackbody radiation shift contributes the largest systematic uncertainty for the individual clocks, at $1\times10^{-19}$, which is limited by the uncertainty in the dc polarizability~\cite{arnold2018blackbody,arnold2019dynamic}. The differential BBR uncertainty is limited by the assessment of the environment temperature which is detailed in the Supplementary Information. Briefly, the assessment bounds the temperature between that of the ambient environment and the maximum temperature of the ion's local environment due to rf heating of the ion trap. The contribution of rf heating is evaluated by temperature measurements on a test trap of identical construction combined with modeling and numerical simulations of the heat transfer. We estimate the temperatures $T_1 = 301.5(1.7)\,$K for Lu-1 and $T_2 = 300.8(1.2)\,$K for Lu-2 which capture, within $1\sigma$ uncertainty, the minimum and maximum temperature bounds. We note it would require an increase in the upper bound temperature by 2 degrees in order to change the BBR uncertainties of the individual systems even at the lowest significant digit reported,  $1\times10^{-20}$. 

Some of the shifts in \tref{tab:sys} are suppressed by the use of long interrogation times: the ac Zeeman shift due to the microwave interrogation pulses, the ac Stark shift due to the 848 nm clock laser, and HA timing errors due to microwave coupling uncertainties. On the other hand, the second-order Doppler shift (SODS) due to thermal motion becomes more significant as the interrogation time is increased due to ion heating. Heating rates for both traps were characterized separately for radial and axial principal axes. At $T_R =5\,$s, the SODS uncertainty is still at the low $10^{-20}$ level, but would become the largest source of uncertainty for significantly longer interrogation times. 

Both the ac magnetic field at the rf trap frequency and the dc electric field gradient are characterized by auxiliary measurements on Lu$^+$, in contrast to previous work that used Ba$^+$~\cite{arnold2020precision}. The ac magnetic field is inferred by measuring an Autler-Townes splitting on the $\ket{^3D_2,6,0}$ to $\ket{^3D_2,5,0}$ transition with microwave spectroscopy. The electric field gradient is inferred from the quadrupole shift measured on the $^3D_1$ microwave transitions, made possible by recent precision measurement of the unperturbed frequencies of the microwave clock transitions~\cite{lee2025quadrupole}. Systematic uncertainty in the quadrupole shift on the HA optical frequency is limited by the uncertainty in the residual quadrupole moment~\cite{zhiqiang2020hyperfine,lee2025quadrupole}.  

At the level of accuracy and precision demonstrated in this work, the shift arising from collisions between the ion and the background gas is a potentially important systematic.  We have investigated this through both a classical analysis similar to that in \cite{hankin2019systematic} and via a quantum treatment given in \cite{vutha2017collisional} for which we get agreement from both descriptions.  We have shown~\cite{mbd_collisions} that the shift can be bounded by the classical Langevin collision rate and a factor that characterizes the decoupling of the ion from the clock laser due to the recoil motion.  Based on collision rates inferred from such decoupling measurements, we infer background pressures of 3.5 and 5.8 nPa and estimate collision shift bounds of $3.9\times10^{-20}$ and $3.2\times10^{-20}$ in the respective chambers (see Methods).

\section*{Discussion}

An ideal optical reference transition has three essential properties~\cite{fortier2026optical}:  a high frequency and long-lived upper state, enabling good short-term stability; insensitivity to external perturbation, enabling high accuracy; and practical operation with accessible laser wavelengths. The $^1S_0\leftrightarrow {^3}D_1$ transition in $^{176}$Lu$^+$ satisfies all of these criteria to a high degree simultaneously, perhaps more than any other reference.  The week-long excited-state lifetime removes any practical limit to interrogation time. The cooling, repumping, and clock wavelengths are all accessible with commercial solid-state and diode laser sources.  The insensitivity to external perturbation has made possible the record-low evaluated uncertainty reported here, even though these experiments are operated at room temperature in relatively basic linear Paul traps without magnetic shielding, which may be readily developed into transportable optical clocks~\cite{grotti2018geodesy} without compromising performance.  Moreover, an improved evaluation of the dc polarizability as proposed in~\cite{barrett2025extrapolation} would immediately reduce the total uncertainty of these existing systems to the level of mid $10^{-20}$, although our measurement precision must advance to take advantage of this capability. These properties in combination make $^{176}$Lu$^+$ a compelling candidate as the international community assesses systems for the redefinition of the SI second.

Frequency comparisons are the means by which clock performance is established in practice. The international roadmap towards redefinition of the SI second~\cite{dimarcq2024roadmap} sets two thresholds: an evaluated systematic uncertainty of $\lesssim 2\times10^{-18}$ and validation through frequency comparisons between institutes demonstrating agreement to an overall uncertainty of $\lesssim 5\times10^{-18}$. Between-institute comparisons at this level remain an outstanding challenge, with only a limited number of within-institute same-species comparisons having reached this threshold.  \tref{tab:comparison} summarizes all reports, to the best of our knowledge, that meet either threshold. Looking beyond these thresholds, differential gravitational redshift between remote sites becomes an unavoidable limitation. Over long baselines, geopotential differences are presently known to no better than a few centimeters in equivalent height, corresponding to a relative frequency uncertainty $\gtrsim 2\times10^{-18}$~\cite{denker2018geodetic,mehlstaubler2018atomic}, well above both the evaluated uncertainty and the verification level of this work. Below $10^{-18}$ testing of uncertainty budgets thus requires comparison of references at the same location, with inter-institute validation at this level only possible via transportable references~\cite{grotti2018geodesy}. Of the standards listed in \tref{tab:comparison}, fewer than half are accompanied by a same-species comparison, and prior to this work none have total uncertainty below $10^{-18}$.

Accuracy reflects reproducibility, for which we are limited in this report by the measurement precision.  But reproducibility alone is insufficient if large systematics are incorrectly evaluated, which is why we have stress tested the quadratic Zeeman shift multiple times.  In contrast, the BBR shift is still too small for practical temperature differences to influence the observed agreement.

\begin{table}[t]
\caption{{\bf Summary of clock accuracy and comparison results.} Values are relative to the optical frequencies of the respective standards. All reports, to the best of our knowledge, which satisfy either of the SI redefinition roadmap~\cite{dimarcq2024roadmap} thresholds of evaluated systematic uncertainty $\lesssim 2\times 10^{-18}$ or same-species comparison with an overall uncertainty $\lesssim 5\times 10^{-18}$. }
\label{tab:comparison}
\begin{tabular}{p{28mm}p{18mm}p{20mm}p{38mm}}
\toprule[0.75pt]
Species \newline (Institute) & Year \newline (Ref) & Systematic \newline Unc. $[10^{-18}]$ & Comparison Difference \newline $[10^{-18}]$ \\
\midrule[0.5pt]
Yb (NIST) & 2018~\cite{mcgrew2018atomic} & 1.4 & $-0.7 \pm (0.5)_\mathrm{stat} \pm (0.8)_\mathrm{sys}$\\
Yb$^+$ (PTB) & 2019~\cite{sanner2019optical} & 2.7 & $+2.8 \pm (2.1)_\mathrm{stat} \pm (3.6)_\mathrm{sys}$ \\
Al$^{+}$ (NIST) &  2019~\cite{brewer2019al+} & 0.94 & -\\
Sr (JILA) & 2019~\cite{bothwell2019jila} & $2.0$ & - \\
Sr (RIKEN) & 2020~\cite{takamoto2020test} &$5.5$ &$-0.3 \pm (3.1)_\mathrm{stat} \pm (3.5)_\mathrm{sys}$\\
Lu$^{+}$ (NUS) & 2023~\cite{zhiqiang2023} & $0.8^\dagger$ & $-2.0 \pm (3.7)_\mathrm{stat} \pm (0.9)_\mathrm{sys}$\\
Sr (JILA) & 2024~\cite{aeppli2024clock} & 0.81 & - \\
Sr$^{+}$ (VTT MIKES) & 2025~\cite{lindvall202588} & $0.79$ & -\\
Al$^{+}$ (NIST) &  2025~\cite{marshall2025high} & $0.55$ & -\\
Ca$^{+}$ (SKLC) & 2026~\cite{zhang2026liquid} & $0.44$ & -\\
Sr (USTC) &2026~\cite{jia2026improved} &0.92 &-\\
Lu$^{+}$ (NUS) & This work & $0.12$ & $-0.01 \pm (0.57)_\mathrm{stat} \pm (0.10)_\mathrm{sys}$ \\
\bottomrule[0.75pt]
\end{tabular}
$^\dagger$Assuming the BBR shift evaluated for 35(10)$\,^\circ\mathrm{C}$ as stated in~\cite{zhiqiang2023}
\end{table}

At the level of precision demonstrated here, a comparison between remote references is necessarily a direct measure of geopotential difference rather than a test of the optical references. This sensitivity resolves height differences at the millimeter scale, beyond the reach of conventional geodetic techniques~\cite{lion2017determination,mehlstaubler2018atomic}. Likewise, for tests of fundamental physics~\cite{safronova2018search} an unexplained frequency difference is evidence of new physics only if the references themselves are above suspicion. Fulfillment of this potential therefore requires rigorous, measurement-based assessment and demonstrated reproducibility of the standards.  Below $10^{-18}$, lutetium stands alone in this regard.


\backmatter
\section*{Methods}

\textbf{Experimental Sequence}
Each interrogation begins only after both ions have been Doppler cooled and prepared in the $\ket{g}$ state with $>99\%$ probability using the conditional state preparation sequence described in~\cite{zhao2025lande}. Both ions are interrogated with the HA-HR sequence shown in Fig.~\ref{fig:scheme}d with the parameters $\tau_L \approx 4\,$ms, $\tau_1=\tau_2\approx12\,$ms, and a total Ramsey time given by $T_R= 3(T+\tau_1+\tau_2)$.  Comparisons of the two independent frequency references are carried out by measuring the parity alternately for $\phi=\pm\frac{\pi}{2}$ and, after $N$ interrogation cycles, steering $\Pi(\frac{\pi}{2})-\Pi(-\frac{\pi}{2})$ to zero by updating $\delta f$. For the comparisons at the same 0.1 mT magnetic field, as is the case for all results shown in Fig.~\ref{fig:results}, the microwave drive frequencies $f_1$ and $f_2$ are identical for both ions and fixed throughout. For comparisons at different magnetic fields, as when evaluating the quadratic Zeeman coefficient $\alpha_z$, the microwave frequencies are necessarily different to compensate quadratic Zeeman shifts. The frequency difference $\delta f$ is set by an acousto-optic modulator (AOM), labelled AOM 2a in Fig.~\ref{fig:scheme}a. Two auxiliary measurements are interleaved with the comparison servo: Rabi spectroscopy of the $\ket{g} \leftrightarrow \ket{8}$ optical transition to ensure the 848 nm clock laser is near resonance and measurement of the $\ket{^3D_1,6,\pm1}$ Zeeman splitting for each ion using microwave spectroscopy. The Zeeman splitting is used to infer the magnetic field, and feedback is applied to shim coils to compensate for any slow field drift, with further details in the Supplementary Information.  HA-HR interrogation time is typically 87\% of the total duty cycle, including the auxiliary measurements.\\

\textbf{Magnetic field stability and servo} 
The short-term stability of the magnetic field is shown in Extended Data Fig.~\ref{fig:mag} (blue points). This was evaluated from a separate experiment measuring a Zeeman splitting with fast servo attack time. Over the range of Ramsey times of interest in this work ($\sim$ 1 to 50 seconds), the magnetic field noise shows flicker instability of approximately 7 nT.  During comparison measurements the magnetic field is steered to the set point of $B_0 =0.1$ mT by a compensation coil with a typical servo attack time of $t_\mathrm{ser} \approx 80$ s using interleaved measurements of the $\ket{^3D_1,6,\pm1}$ Zeeman splitting. The orange points in Extended Data Fig.~\ref{fig:mag} show the field instability inferred from the in-loop measurements of the $\ket{^3D_1,6,\pm1}$ Zeeman splitting which averages down $\propto (t_\mathrm{ser}/\tau)$ for reasons explained in the Supplementary Information.  The true instability of magnetic field is limited by both projection noise and deadtime  $\propto (t_\mathrm{ser}/\tau)^{1/2}$, which was verified in a test run using interleaved out-of-loop measurements of the $\ket{^3D_1,8,\pm1}$ Zeeman splitting (Extended Data Fig.~\ref{fig:mag} green points). This out-of-loop stability is taken as an upper bound on the magnetic field instability with compensation engaged. The magnetic field contributes uncertainty via the quadratic Zeeman shift of $\frac{\delta \nu_\mathrm{QZ}}{\nu_0} = 2\alpha_z B_0 \delta B(\tau) \approx 2\times10^{-19} \cdot (\tau/\mathrm{s})^{-1/2}$ beyond the servo attack time. \\

\textbf{Detection and background gas collisions}
In order to detect background collisions as much as possible and ensure the ions are sufficiently re-cooled before the next experiment cycle, we use the detection sequence shown in Extended Data Fig.~\ref{fig:collision}{a} at the end of every Ramsey experiment. The intervals $d_i$ represent adaptive Bayesian state detection~\cite{myerson2008high} by 646 nm fluorescence. If the ion is detected dark in the initial detection $d_0$ after the Ramsey sequence, which is $\sim$50\% of events, then three attempts are made to shelve on the 848 nm clock transition and detect. The variable 848 nm clock pulse areas are tailored to maximize population transfer even for higher thermally-occupied vibrational $n$ states. The probability of at least one successful shelving is $>99.5\%$ for the expected thermal distribution accounting for ion heating during the Ramsey dark time $T_R$. If the ion is not detected bright on any of $d_{0..3}$ then it is assumed that a collision has occurred with sufficient energy transfer to significantly reduce either (i) the coupling on the 848 nm transition or (ii) the 646 nm fluorescence rate. During clock comparison servo operation, if a collision is detected in this way on either Lu-1 or Lu-2, that interrogation cycle is considered invalid and repeated. 

If a collision is detected ($d_{0..3}$ all dark), the additional ending sequence shown in Extended Data Fig.~\ref{fig:collision}{a} is applied. First a repump pulse (350, 622, and 895 nm) and another detection $d_4$ is attempted. If bright, this is attributed to a lower energy transfer collision of type (i): sufficient to reduce efficiency of shelving on the clock transition but not detection. If dark, then the collision is of type (ii), energetic enough to disrupt detection. In either case, a cycle of cooling and fluorescence detection with a high threshold, $b_i$, is repeated until the ion is confirmed bright to ensure the ion is effectively cooled for the next experiment cycle. Extended Data Fig.~\ref{fig:collision}{b} shows the probability of outcomes (i) and (ii) as a function of $T_R$ from which we extract rates $\Gamma^{(i)}$ and $\Gamma^{(ii)}$. To estimate the detectable collision rate $\Gamma$, we assume $\Gamma^{(i)} = \frac{1-P_d}{2} \Gamma$ and $\Gamma^{(ii)} = P_d \Gamma$ where $P_d$ is the probability a collision interferes with bright detection, resulting in outcome (ii). Of the lower energy transfer collisions at rate $(1-P_d)\Gamma$, we assume half are not detected because the ion was in $^3D_1$ at the end of the Ramsey experiment and half are detected as outcome (i). We infer the collision rates $\Gamma = 1.9\times 10^{-3}\,\mathrm{s}^{-1}$ for Lu-1 and $5.9\times 10^{-3}\,\mathrm{s}^{-1}$ for Lu-2. 

After the comparison measurement campaign, additional experiments were performed to estimate the collision rate without the complications introduced by Ramsey spectroscopy. The ions were prepared in the $^1S_0$ ground state, and after a delay of 5 s, an attempt is made to reshelve and detect using a similar sequence as shown in Extended Data Fig.~\ref{fig:collision}{a}.  From the probability that all of multiple shelving attempts on the clock transition fail, we directly infer the detectable collision rates $\Gamma = 2.9(3)\times 10^{-3}\,\mathrm{s}^{-1}$ for Lu-1 and $6.5(6)\times 10^{-3}\,\mathrm{s}^{-1}$ for Lu-2, in reasonable agreement with the rates inferred from the measurement campaign data. 

Our extensive analysis of collision shifts in ion-based optical clocks is given in~\cite{mbd_collisions}. The collision shift is evaluated with respect to a Langevin collision rate $\Gamma_L$, defined as the rate of collisions below a critical impact parameter which result in an inward-spiraling trajectory.  We calculate the overestimate of the Langevin collision rate using \cite[Eq~(31)]{mbd_collisions}, and with the cutoff velocity given by the 50\% threshold of the Ramsey suppression factor as defined in \cite[section 2]{mbd_collisions}. Using $\omega\sim1\,$MHz for Lu-1 and $\omega\sim500\,$kHz for Lu-2 yields $\Gamma_L = 1.3\times 10^{-3}\,\mathrm{s}^{-1}$ and $2.1\times 10^{-3}\,\mathrm{s}^{-1}$ for the respective Langevin rates, corresponding to 3.5 and 5.8 nPa background pressures of molecular hydrogen at 300$\,$K and consistent with pressure gauge readings. An inverted magnetron gauge on Lu-1 reads 4 nPa, and the ion pumps on both chambers read `low pressure' ($<13\,$nPa) at the limit of the sensitivity of the ion pump controllers. With $\kappa=1$ in \cite[Eq~(60)]{mbd_collisions}, we estimate collision shift bounds of $3.9\times10^{-20}$ and $3.2\times10^{-20}$.\\

\textbf{Blackbody Radiation}
The differential dynamic polarizability, $\Delta\alpha(\omega)$, for the 848-nm clock transition in $^{176}$Lu$^+$ has been well characterized~\cite{arnold2018blackbody,arnold2019dynamic}. The BBR shift is given by 
\begin{equation}
\frac{\delta \nu_\mathrm{BBR}}{\nu_0}=-4.90\times 10^{-19}\bar{T}^4 (1+1.77 \bar{T}^2),
\end{equation}
where $\bar{T}\equiv T/T_0$ and $T_0 = 300$ K. Over the practical temperature range of 270-330\,K, the uncertainty contribution from $\Delta\alpha(\omega)$ is well approximated by $9.8\times 10^{-20} \bar{T}^4$~\cite{arnold2019dynamic}.
A detailed temperature assessment is given in the Supplementary Information in which we bound the minimum and maximum temperatures to $[299.8, 303.2]\,$K for Lu-1 and $[299.6,301.9]\,$K for Lu-2.\\

\textbf{Trap Secular Frequencies}
The Lu-1 trap used in previous work~\cite{zhiqiang2023} had an unusually high heating rate which has been resolved after replacing the ion trap with one of the same design~\cite{zhao2025lande}.  Additionally, the helical resonators for both Lu-1 and Lu-2 were changed to reduce the rf drive frequency $\Omega_\mathrm{rf}$ and obtain higher trap confinement without increased rf power. For 0.25 W of rf power at $\Omega_\mathrm{rf} = 2\pi \times 9.4$ and 11.2 MHz, we obtain secular trapping frequencies of (207, 1063, 1134) kHz and (138, 492, 543) kHz in Lu-1 and Lu-2 respectively, where the weakest confinement corresponds to the axial direction. The combined effect of reduced heating rates and the smaller Lamb Dicke parameters resulting from the increased radial confinement means ion heating is no longer a significant limitation to the interrogation time. The simulated contrast loss shown in Fig.~\ref{fig:results}{a} is evaluated by averaging the final optical pulses over a thermal distribution of radial mode occupation accounting for the heating during the Ramsey time.\\

\textbf{Thermal Second-Order Doppler}
Thermal motion gives rise to a second-order Doppler shift (SODS) given by
\begin{equation}
\frac{\delta\nu_\mathrm{SODS}}{\nu_0} = -\frac{\langle v^2 \rangle}{2 c^2} = -\frac{k_B}{2m c^2} \sum_{i} \left(\kappa_i T_i\right) \label{eq:thermal}
\end{equation}
where $T_i$ are the average temperatures during the Ramsey interrogation for the axial ($i=1$) and radial ($i=2,3$) principal axes, respectively. The geometry factor $\kappa_2=\kappa_3\approx2$ accounts for equal contributions from the secular motion and intrinsic micromotion (IMM) for the two radial principal axes, which is valid when the confinement is predominantly ponderomotive~\cite{berkeland1998minimization}. Axial confinement is predominantly due to the static potential and so only secular motion contributes ($\kappa_1\approx1$). 

The initial temperatures and heating rates are determined by measuring the temperature immediately after Doppler cooling and after fixed delay time. The axial temperature is measured by fitting the thermal dephasing of Rabi flopping on the 804 nm E2 transition. Although the 804 nm laser wave vector is at $45^\circ$ with respect to the ion trap axis and has a projection onto all principal axes, the radial modes have sufficiently low thermal occupation $\bar{n}$ and a small Lamb-Dicke parameter $\eta$ that they contribute negligibly to the thermal dephasing. From the weighted mean of three measurements over the course of the campaign, we find the initial temperatures $T_{\mathrm{1},0} = 190.5(5.9)\,\mu$K for Lu-1 and $189.6(4.7)\,\mu$K for Lu-2 and heating rates $\frac{\mathrm{d}T_\mathrm{1}}{\mathrm{d}t} = 73(14)\,\mu$K/s for Lu-1 and $119(17)\,\mu$K/s for Lu-2, summarized in Extended Data Fig.~\ref{fig:axtherm}. The average temperature during the Ramsey experiment for each mode is given by $T_i = T_{\mathrm{i},0} + \frac{\mathrm{d}T_\mathrm{i}}{\mathrm{d}t}\frac{T_\mathrm{R}}{2}$.

The radial temperatures are measured by spectroscopy on the secular motion sidebands using the 804 nm E2 transition. The average thermal occupation $\bar{n}$ of one of the radial modes is extracted from the ratio of population transferred on the red and blue sidebands~\cite{turchette2000}:
\begin{equation}
\frac{P_\mathrm{red}(t)}{P_\mathrm{blue}(t)} = \frac{\bar{n}}{\bar{n}+1}.
\end{equation}
For both traps, the initial $\bar{n}$ measurements on both radial modes are consistent with the temperature $150(30)\,\mu$K, which is also consistent with the initial axial temperatures to within statistical uncertainty and corresponds to approximately three times the Doppler cooling limit. The sideband ratio method is most sensitive for low $\bar{n}$, so for measuring the radial heating rates we first apply Zeeman-degenerate Raman sideband cooling~\cite{qichen2025} to prepare in the motional ground state and then measure the sideband ratio after some delay. Extended Data Fig.~\ref{fig:radtherm} shows the results of radial heating measurements in both chambers, which yield heating rates $[\frac{\mathrm{d}T_\mathrm{2}}{\mathrm{d}t}, \frac{\mathrm{d}T_\mathrm{3}}{\mathrm{d}t}] = [110(15), 29.1(2.9)]\,\mu$K/s for Lu-1 and $[85.5(5.5), 70.4(5.5)]\,\mu$K/s for Lu-2.  
 
The \eref{eq:thermal} is the SODS for a thermal state in the high temperature limit, $\bar{n}_i\approx \frac{k_B T_i}{\hbar\omega_i}$, and does not account for the zero-point fluctuations for the quantum ground state for both secular and micromotion~\cite{massdefect2022,propertime2026}, which contribute additional SODS of $-\sum_{i} \kappa_i\left(\frac{ \hbar \omega_i}{4 m c^2} \right)$. This term contributes only a 6\% correction to the total SODS for Lu-1 but has been included for completeness.
  
 We estimate the total SODS for a Ramsey time $T_\mathrm{R}$ to be:
 \begin{align*}
 \mathrm{(Lu-1)}~ ~ \frac{\delta\nu_\mathrm{SODS}}{\nu_0} =& -2.4(2)\times10^{-19} + T_\mathrm{R}\cdot [-4.6(4)\times 10^{-20} \mathrm{s}^{-1}]\\
 \mathrm{(Lu-2)}~ ~\frac{\delta\nu_\mathrm{SODS}}{\nu_0} =& -2.4(2)\times10^{-19} + T_\mathrm{R}\cdot [-5.7(3)\times 10^{-20}  \mathrm{s}^{-1}]\\
 \end{align*}

 \textbf{Excess Micromotion}
 The micromotion shift has two components: a SODS due to oscillatory motion and the ac-Stark shift due to the rf electric field. Owing to the very low differential polarizability of the clock transition, the ac-Stark contribution is over two orders of magnitude smaller than the SODS component and therefore negligible. The micromotion amplitude is evaluated using phase-modulated sideband spectroscopy on the 804-nm E2 clock transition~\cite{arnold2024enhanced}. The excess micromotion (EMM) second-order Doppler shift is given by  
\begin{equation}
\frac{\delta \nu_\mathrm{EMM}}{\nu_0}=-\left(\frac{\Omega_\mathrm{rf}}{2 c k_\mathrm{804}}\right)^2 |\beta|^2 
\end{equation}
where $k_\mathrm{804}$ is the wave number of the 804-nm clock transition and $\beta$ is the modulation depth. Since phase-modulated sideband spectroscopy is sensitive to the phase of the micromotion, we distinguish between two quadrature components $\beta=\beta_m+i \beta_p$, where $\beta_m$ is the excess micromotion due to displacement by stray fields, and $\beta_p$ is due to a phase shift between rf-electrodes which we are not able to compensate. Intrinsic micromotion (IMM) is accounted for separately with the thermal second-order Doppler shift.

Following the procedure described in \cite{arnold2024enhanced}, we measure the modulation depth in three orthogonal directions to evaluate $\beta^2_m = \sum_i \beta^2_{m,i}$.  The results of all measurements of $\beta_{m,i}$ taken for both ion traps are shown in Extended Data Fig.~\ref{fig:emm}a-b, where the open circles are measured before compensating the dc stray field and closed circles immediately after. To estimate the EMM shift for each comparison measurement, we linearly interpolate the $\beta_{m,i}$ measurements, assuming linear growth of the dc stray field between measurements. Extended Data Fig.~\ref{fig:emm}c shows the EMM shifts evaluated from the time average of $\beta_m^2$ over each comparison interval. We estimate an average relative shift of $-1.6(1.3)\times10^{-20}$ for Lu-1 and $-1.4(0.4)\times10^{-20}$ for Lu-2. 

As reported in~\cite{arnold2024enhanced}, $\beta_p$ is negligibly small for Lu-1. For Lu-2, the three orthogonal components of $\beta_p$ are measured to be $[2.22(13),1.43(12),1.92(13)] \times 10^{-2}$, corresponding to an EMM shift of $-2.42(19)\times10^{-19}$. $\beta_p$ was remeasured at the end of the campaign and found to be in statistical agreement with the original measurement. \\

 \textbf{AC Zeeman (rf)}
Time-varying magnetic fields give rise to an ac-Zeeman shift, with the dominant contributions coming from currents in the electrodes driven by the rf–trapping potential. The contribution depends on the component of the rf-magnetic field perpendicular to the applied dc field~\cite{gan2018oscillating}. The clock shift given by
\begin{equation}
\delta \nu_{\mathrm{rf}} = \alpha_\perp \langle B_\perp^2\rangle 
\end{equation}
where $\sqrt{ \langle B_\perp^2\rangle }$ is the root-mean-square rf magnetic field amplitude perpendicular to the dc field and $\alpha_\perp = 0.20\,\mathrm{mHz}/\mu\mathrm{T}^2$ is the sensitivity coefficient after $^3D_1$ hyperfine averaging~\cite{gan2018oscillating}. 

While in previous work $B_\perp$ was measured via an Autler-Townes splitting on the Ba$^+$ clock transition~\cite{arnold2020precision}, here we use an Autler-Townes splitting on the Lu$^+$ $^3D_2$ $\ket{6,0}$ to $\ket{5,0}$ microwave transition. The $^3D_2$ $F=5$ Land\'e g-factor, $g_5 = -0.385 757 50 (19)$~\cite{zhao2025lande}, is the largest of all $^3D_1$ and  $^3D_2$ $F$ manifolds and for $\Omega_\mathrm{rf}\sim2\pi\times10\,$MHz the linear Zeeman splitting can be brought into resonance with $\Omega_\mathrm{rf}$ at experimentally accessible magnetic fields of $B\lesssim 2\,$mT. 
 
The experiment measurement procedure is as follows. Neodymium permanent magnets are used to bias the magnetic field to the required 1.7 mT (2.0 mT) for Lu-1 (Lu-2) respectively. Three pairs of coils are used to fine tune the magnetic field amplitude and optimize the orientation for $\ket{^3D_1,7,0}$ state preparation by aligning to a $\pi$-polarized 646-nm laser. This ensures the dc magnetic field is aligned in the same direction as during the comparison measurements.   From $\ket{^3D_1,7,0}$, we transfer sequentially to the $\ket{^3D_1,8,0}$ state using a microwave $\pi$ pulse and then the $\ket{^3D_2,6,0}$ state using an 804-nm/848-nm Raman pair. We observe the Autler-Townes splitting by interrogating the ${^3D_2}$ $\ket{6,0}$ to $\ket{5,0}$ microwave transition with a $\pi$ pulse for a range of detunings, after which the remaining $\ket{^3D_2,6,0}$ population is reshelved to $\ket{^3D_1}$ by Raman transfer for detection. 

Following the treatment in~\cite{gan2018oscillating}, the effective Hamiltonian in the rotating wave approximation describing the microwave interrogation is
\begin{equation}
    H  = \frac{\hbar}{2}\begin{pmatrix}
        -2\Delta_M & \Omega_M & 0 & 0 \\
        \Omega_M & 0 & \Omega_B & \Omega_B \\
        0 & \Omega_B & 2\Delta_1 & 0 \\
        0 & \Omega_B & 0  & -2\Delta_{-1}\label{eqn:AT_H}
    \end{pmatrix}
\end{equation}
where $\Omega_M$ is the microwave coupling, $\Delta_M$ is the microwave detuning,  $\Omega_B = \frac{1}{\hbar}\frac{\sqrt{15}}{2} g_5 \mu_B B_\perp $ is the coupling between $\ket{^3D_2,5,0}$ and  $\ket{5,\pm1}$ states due to the transverse oscillating magnetic field at frequency $\Omega_\mathrm{rf}$, and 
\begin{equation}
  \Delta_{\pm1} = \Omega_\mathrm{rf} \pm \omega_{\pm 1}=  \Omega_\mathrm{rf} +g_5\mu_BB/\hbar \pm \Delta\alpha B^2 \label{eqn:AT_detuning}
 \end{equation}
 is the detuning of the rf field from the $\ket{5,0}$ to $\ket{5,\pm1}$ Zeeman splittings,  $\omega_{\pm 1}$. Degeneracy of $\omega_{\pm 1}$ is lifted by the differential quadratic Zeeman shift for which $\Delta\alpha \approx 1.23\,$kHz/mT$^2$. 
 
 Extended Data Fig.~\ref{fig:autler}{a} shows the simulated spectrum of the Autler-Townes splittings for the operating parameters of Lu-1 and $\Omega_B = 2\pi \times 3\,$kHz. For both Lu-1 and Lu-2, several spectra were taken at a fixed magnetic field in the vicinity of the Autler-Townes splitting, of which three from Lu-1 are shown in Extended Data Fig.~\ref{fig:autler}{b-d}. We fit the measured data to simulated spectra to obtain $\Omega_B$.  Taking the mean of the $\Omega_B$ fit values, with the standard deviation of fit values as the uncertainty, we find $\sqrt{ \langle B_\perp^2\rangle }$ to be 0.220(5) $\mu$T and 0.303(4) $\mu$T for Lu-1 and Lu-2, respectively. The rf-drive voltage to the trap is monitored over the campaign and for both Lu-1 and Lu-2 varied by less than 0.5\%. \\

 \textbf{AC Zeeman (microwave)}
When applying the microwave fields during the Ramsey sequence for hyperfine averaging, there is a probe-induced shift due to the $\sigma_\pm$ polarization components off-resonantly coupling to $m=\pm 1$ Zeeman states. Evaluation of this shift is discussed in detail in the Supplementary Material of~\cite{zhiqiang2023}. The total microwave ac-Zeeman shift is given by 
\begin{equation}
\label{shift}
\delta\nu_\mu=\frac{(\Delta_{1,7}+\Delta_{1,8})\tau_1+(\Delta_{2,6}+\Delta_{2,7})\tau_2}{T_\mathrm{R}},
\end{equation}
where $\Delta_{k,F}$ is the shift of $\ket{F,m=0}$ when microwave coupling $\Omega_k$ is on. The microwave field polarizations are characterized by measuring the $\pi$ times $\tau_{kq}$ for the field $k$ and polarization $q$, at fixed microwave power, from which the shifts $\Delta_{k,F}$ are evaluated as:

\begin{align*}
\Delta_{1,8} &= -\frac{7}{9}\frac{\Omega_{1}^2}{4\omega_7}
\left[
        \left(\frac{\tau_{10}}{\tau_{1+}}\right)^2
        - 
        \left(\frac{\tau_{10}}{\tau_{1-}}\right)^2
\right] \\[2ex]
\Delta_{1,7} &= \frac{\Omega_{1}^2}{4\omega_8}
\left[
        \left(\frac{\tau_{10}}{\tau_{1+}}\right)^2
        -
        \left(\frac{\tau_{10}}{\tau_{1-}}\right)^2
\right] \\[2ex]
\Delta_{2,7} &= \frac{\Omega_{2}^2}{4\omega_6}
\left[
        \left(\frac{\tau_{20}}{\tau_{2+}}\right)^2
        -
        \left(\frac{\tau_{20}}{\tau_{2-}}\right)^2
\right] \\[2ex]
\Delta_{2,6} &= \frac{4}{3}\frac{\Omega_{2}^2}{4\omega_7}
\left[
        \left(\frac{\tau_{20}}{\tau_{2+}}\right)^2
        -
        \left(\frac{\tau_{20}}{\tau_{2-}}\right)^2
\right]
\end{align*}
where $\omega_F>0$ is the Zeeman splitting for the hyperfine level $F$. 

This shift is suppressed by ensuring $\tau_k \ll T_\mathrm{R}$, balancing the circular polarization components ($\tau_{k+}\approx \tau_{k-}$), and maximizing the $\pi$ coupling ($\tau_{k0} \ll \tau_{k\pm}$). All microwave horns are mounted on rotation mounts and are tuned, through a combination of translation and rotation, to the condition $\tau_{k+}\approx \tau_{k-}$ as much as possible at the start of the campaign. All $\tau_{kq}$ were characterized at three points in the campaign, and the evaluated clock shifts were consistently below $1\times10^{-20}$.\\

 \textbf{Quadrupole Shift}
 The hyperfine-averaged residual quadrupole moment has been reported as $\tilde \Theta = -2.54(0.25)\times10^{-4}\,e a_0^2$ \cite{zhiqiang2020hyperfine}, which has been independently confirmed by an alternate method~\cite{lee2025quadrupole}. The residual quadrupole shift, $\delta\tilde{\nu}_Q$, is evaluated by measuring the quadrupole shift, $\delta \nu_{Q,1}$, on the $^3\mathrm{D}_1$ $\ket{7,0}$ to $\ket{8,0}$ microwave clock transition, which are related by
 \begin{equation}
\delta\tilde{\nu}_Q = \frac{\tilde \Theta}{\frac{8}{5}\Theta(^3\mathrm{D}_1)} \delta \nu_{Q,1},
\end{equation}
 where $\Theta(^3\mathrm{D}_1) = 0.63862(74)\,ea_0^2$ \cite{kaewuam2020precision}. 
 We measure the microwave transition frequencies for both traps by microwave Ramsey spectroscopy with a 10 s interrogation time. The quadrupole shifts $\delta \nu_{Q,1}$ are inferred using the unperturbed microwave frequencies~\cite{lee2025quadrupole} and accounting for the second-order Zeeman shifts. The residual quadrupole shifts on Lu-1 and Lu-2 are evaluated to be $3.3(3) \times 10^{-20}$ and $1.35(14) \times 10^{-19}$ respectively. Given the accuracy of the measured unperturbed microwave frequencies~\cite{lee2025quadrupole}, the quadrupole shift may be easily suppressed further by setting the magnetic field angle so as to null the microwave quadrupole shifts as much as is required.\\

\textbf{Microwave Coupling Errors}
Incorrect microwave pulse areas due to uncertainty in the respective couplings give rise to timing errors in HA-HR spectroscopy. As derived in the Supplementary Material of~\cite{zhiqiang2023}, the shifts due to microwave coupling errors are bound by
 \begin{equation}
 \delta\nu_{k} \approx -\frac{1}{T_\mathrm{R}}\left(\left(\frac{\Delta_k \tau_k}{\pi}\right)^2+\left(\frac{\pi q_k}{2}\right)^2\right)\label{eq:eerr}
 \end{equation}
 where $\Delta_k$ is the microwave detuning, $\tau_k$ is the microwave $\pi$ pulse duration, and $q_k$ is the fractional error in the microwave coupling for the $k={1,2}$ microwave transition.
 
These shifts are suppressed for Ramsey times $T_\mathrm{R}\gg \tau_L,\tau_k$ as is the case for the comparison experiments reported here. The magnetic field was actively steered to a fixed value in both chambers so the microwave detunings $\Delta_k$ are stable and known precisely from the quadrupole shift assessment. The microwave couplings were measured at the beginning, middle, and end of the campaign with $\sim0.2\%$ measurement precision and found to deviate by less than 0.5\%, with the exception of $\Omega_2$ on Lu-2, which drifted by 1.5\% from the beginning to the end of the campaign. Extended Data Fig.~\ref{fig:uwcouplingerror}, for example, shows this shift evaluated for the worst case of $q_2=0.015$ as a function of the detuning $\Delta_2$. Although the $\Delta_k$ are precisely known at the $<10$ mHz level, which in principle allows for a more accurate estimation of this shift, we take a conservative approach and use the full width of the bounding envelope, given by second term in \eref{eq:eerr}, as the uncertainty.   \\

\textbf{AC Stark}
For an optical $\pi$ time of $\tau_L = \frac{\pi}{\Omega_L} = 4$ ms, the ac Stark shift on the $\ket{g}$ to $\ket{8}$ optical transition when interrogating with the 848-nm laser is approximately $\Delta_S = 2\pi\times 25$ Hz. In the absence of additional effects, both the hyper-Ramsey~\cite{yudin2010hyper} sequence, Extended Data Fig.~\ref{fig:hyperramsey}{a}, which was used in previous work~\cite{zhiqiang2023}, and time-reversed hyper-Ramsey sequence, Extended Data Fig.~\ref{fig:hyperramsey}{b}, which is used in this work, equivalently suppress the resulting clock shift to $\frac{2}{\pi T_R}\left(\frac{\Delta}{\Omega_L}\right)^3$, in Hz, where $\Delta = \Delta_\mathrm{SP} - \Delta_\mathrm{S}$ is the error in the frequency step, $\Delta_\mathrm{SP}$, applied during the optical interrogation pulses. To set the laser frequency step, $\Delta_\mathrm{SP}$, the ac-Stark shift was measured to better than 1\% for both Lu-1 and Lu-2 at the start of the campaign by interleaved self-comparison of Rabi and hyper-Ramsey spectroscopy. The value of the $\Delta_\mathrm{SP}$ was fixed throughout the campaign, and we bound the uncertainty in $\Delta$ to 2\% of $|\Delta_S|$ for Lu-1 and 5\% of $|\Delta_S|$ for Lu-2 based on the maximum observed variation in the optical couplings, which were measured at the beginning, middle, and end of the campaign with measurement precision of  $\sim0.2\%$.
 
When including the effects of ion heating, hyper-Ramsey schemes present a weak linear dependence on $\Delta$~\cite{kuznetsov2019effect}.  This linear dependence is evaluated by simulation with the results for the typical operating parameters of Lu-2 shown in Extended Data Fig.~\ref{fig:hyperramsey}{c}.  It is noted that the linear slope has the opposite sign for HR and HR reverse schemes, and that a hybrid error signal constructed by averaging the two schemes  cancels the thermal effect, restoring the cubic dependence on $\Delta$. The idea of combining sequences to generate a more robust error signal is similar to modified hyper-Ramsey (MHR)~\cite{hobson2016modified} and generalized hyper-Ramsey (GHR)~\cite{zanon2017universal}, but unlike those schemes for which ion heating results in either an offset or increased sensitivity compared to the basic HR scheme~\cite{kuznetsov2019effect}, this simple hybrid combination of HR and HR reverse heavily suppresses the linear dependence on ion heating, at least in the regime $\tau_L \ll T_R$ where heating during the optical pulse is negligible compared to during the dark time $T_R$. 

For the experiments reported here, the shift was already sufficiently small that only HR reverse was used. From the measured radial heating rates, uncertainty in $\Delta$, and the linear dependence determined from simulation, we estimate a total uncertainty from the ac Stark effect to be $3.8\times10^{-21}$ for Lu-1 and $2.5\times10^{-20}$ for Lu-2 for $T_R=5\,$s.\\

 \textbf{AC Quadrupole (rf)}
 The oscillating rf quadrupole field couples off-resonantly to hyperfine transitions giving rise to an ac quadrupole shift that is not cancelled by hyperfine averaging~\cite{arnold2019oscillating}. We estimate this shift by assuming an ideal linear Paul trap with rf-potential of the form $\Phi(x,y,z) = \epsilon(x^2 -y^2)$ in the principal-axis frame, where the electric field gradient $\epsilon = \frac{m\Omega_\mathrm{rf} \omega_r}{e \sqrt{2}}$ is determined by the radial pseudo-potential confinement frequency $\omega_r \approx 2 \pi \times 1100\,$kHz for Lu-1 and $510\,$kHz for Lu-2. The magnetic field in both chambers is aligned to approximately 33(3)$^\circ$ with respect to the ion trap axis ($z$). Under these conditions we evaluate~\cite{arnold2019oscillating} a relative clock shift of the HA frequency of $-2.0\times10^{-21}$ for Lu-1 and $-5.7\times 10^{-22}$ for Lu-2.\\

\textbf{Differential AOM chirp}
As shown schematically in Fig.~\ref{fig:scheme}a, the differential path length is actively stabilized with reference to retroreflecting mirrors below the respective experimental chambers which are near to the table surface and approximately 20 cm below the trapped ions.  The differential phase stability is characterized by out-of-loop measurement of the optical phase with the clock beams directed to a common beam splitter instead of the ions, requiring $\approx2\,$m of additional unstabilized optical path length. When simultaneously switching on AOMs 1a and 2a, as for the Ramsey pulses in the comparison interrogation sequence, a differential phase chirp is induced by the lock circuitry as shown in Extended Data Fig.~\ref{fig:fods}a. This is well modeled by a damped harmonic oscillation with the fit parameters shown in Extended Data Fig.~\ref{fig:fods}a. By straightforward extension of the analysis given in~\cite{falke2012delivering}, we evaluate a $-4.8(8)\times10^{-21}$ systematic shift to the difference frequency between Lu-1 and Lu-2 for $T_R=5\,$s.\\

\textbf{First Order Doppler}
We consider two sources of first order Doppler shift (FODS): drift of the ion with respect to the trap along the clock probe (vertical) direction and differential drifts of the optical phase.

The measured micromotion modulation depth $\beta_{m,3}$ can be directly related to the displacement of the ion from the trap rf null along the clock interrogation direction~\cite{arnold2024enhanced}.   For small modulation depth, the relationship is linear with scale factors of  approximately 0.77$\,\mu$m/$\beta_{m,3}$ for Lu-1 and 2.0$\,\mu$m/$\beta_{m,3}$ for Lu-2. Extended Data Fig.~\ref{fig:emm}d shows the evaluated FODS shift for each comparison measurement where the average velocity is estimated from the inferred displacement between the EMM measurements.  The weighted FODS for the entire campaign is estimated to be $2.4(3.5)\times10^{-23}$ for Lu-1, $-0.3(2.0)\times10^{-22}$ for Lu-2 and $0.5(2.0)\times10^{-22}$ for the difference.

Using the same measurement setup as used to evaluate AOM chirp, we measure the out-of-loop differential phase for long durations for which we observed an instability asymptote of $3.0\times10^{-16}\,(\tau/\mathrm{s})^{-1}$ as shown in Extended Data Fig.~\ref{fig:fods}b (blue). For typical servo operation at 87\% interrogation duty cycle, we evaluate a contribution of $2.0\times10^{-17}\,(\tau/\mathrm{s})^{-1/2}$ (orange). This is well below the comparison instability on all timescales. When operating as a single clock the path length is actively stabilized from the laser to reference mirror instead of differentially between chambers, but can be expected to contribute at the same level, or lower due to the shorter unstabilized path. \\

\textbf{RF synthesis}
The rf synthesizers used for the AOMs are based on the AD9912 chip and have $\sim7\,\mu$Hz resolution. This contributes $\sim1\times10^{-20}$ fractional uncertainty, half of the minimum step size, to the comparison servo. The synthesized microwave frequencies, $f_1$ and $f_2$ were identical and generated from the same sources for Lu-1 and Lu-2, set to the nearest 1 mHz.  All synthesizers are referenced to a common hydrogen maser which is calibrated to $\lesssim 2 \times 10^{-15}$ at the start of the measurement campaign~\cite{zhang2025absolute}. For the comparison measurement any error in maser accuracy is common mode and contributes no uncertainty to the difference. 

However, we note that when realizing a Lu$^+$ standard via HA-HR spectroscopy as $\nu_0 = f_L + \frac{1}{3}(2f_1+f_2)$, the accuracy of the synthesized microwaves $f_k$ is also not a limitation to optical clock accuracy because the servo action steers the laser to compensate any errors in the microwave fields such that this specific linear combination of the fields is the HA frequency. The HA-HR servo is sensitive to the instability of both the laser and microwave fields at the same level, in Hz.  In practice this places a more relaxed requirement on the microwaves as compared to the laser, e.g. for a laser with fractional instability $\frac{\delta f_L}{f_L} \sim 10^{-16}$, the microwaves need only a fractional stability better than $\frac{\delta f_k}{f_k} \lesssim \frac{\delta f_L}{f_k}\sim10^{-12}$ to not limit the clock instability.  \\

\textbf{Gravitational redshift}
The differential redshift between the ions is given by $\frac{g \delta h}{c^2}$ where $g\approx9.776\,\mathrm{m} \mathrm{s}^{-2}$ is the local gravitational acceleration and $\delta h = h_1 -h_2$ is the height difference of the ions. We measure the heights of the ions relative to a laser leveling assembly fixed to the optical table which is precisely aligned to each ion by the same imaging optics used for state detection. We determine the ion heights $h_1 = 7.44(10)\,\mathrm{mm}$ and $h_2 = 11.41(10)\,\mathrm{mm}$. Measurements of the table leveling by a precision spirit level with 0.02 mm/m resolution are limited by the table surface flatness but bound the table tilt to $<$0.2 mm/m. The ion traps are separated by 1.2 m, and table leveling contributes the dominant uncertainty of 240 $\mu$m. The height difference of the ions is evaluated to be $\delta h = -3.97(27)$ mm, corresponding to a $-4.31(29)\times10^{-19}$ differential shift.\\

\textbf{Clock stability}
The stability of a single-ion optical clock is quantum-projection-noise limited at the longest achievable interrogation time, which is set by either the optical coherence of the local oscillator or the atomic coherence, whichever is shorter. State-of-the-art cavity-stabilized lasers reach fractional instabilities in the low $10^{-17}$ at one second~\cite{oelker2019demonstration,lee2026frequency}, and have enabled demonstrated single-ion clock instabilities as low as $3.5\times10^{-16}(\tau/\mathrm{s})^{-1/2}$~\cite{marshall2025high}. Although our laboratory lacks a local oscillator of this quality with which to challenge ion clock stability at this level directly, our comparison does probe the limits set by atomic coherence, allowing us to infer the technical limits to interrogation time independent of any clock laser. Comparison of two references by correlation spectroscopy carries a factor $\sqrt{2}$ higher instability than a comparison of independent clocks, and a factor of two higher than a single standalone clock, for equal interrogation time. Insofar as the present comparison is limited by uncorrelated magnetic field noise, these systems, when combined with state-of-the-art laser stabilization, would support a single-clock instability of  $\sim2.4\times10^{-16}(\tau/\mathrm{s})^{-1/2}$. Because the $^3D_1$ lifetime imposes no practical limit, substantially longer interrogation times remain possible, contingent on overcoming magnetic field noise, ion trap heating, and background gas collisions. Reducing the dc magnetic field lowers the linear sensitivity to field noise at the cost of smaller Zeeman splittings; nevertheless, operation at 25$\,\mu$T is achievable with modest changes to our laser configuration. By our assessment, this would permit interrogation times up to 30$\,$s in the present system. Even longer coherence times can be anticipated with the use of magnetic shielding and cryogenics, albeit at the cost of increased experimental complexity.\\


\bmhead{Author contributions}

K.J.A and M.D.K.L. performed the experiments and independently analyzed the data.  K.J.A prepared the manuscript.  M.D.B. supervised the project. K.J.A, M.D.K.L., Q.Z, Q.Q., Z.Z., N.J. and M.D.B. contributed to the design of the experiment, discussion of results, and revision of the manuscript. \\

\bmhead{Funding}

This project was supported by the National Research Foundation, Singapore through the National Quantum Office, hosted in A*STAR, under its Quantum Engineering Programme 3.0 Funding Initiative (W25Q3D0007) and under its Centre for Quantum Technologies Funding Initiative (S24Q2d0009).\\

\bmhead{Competing interests}
The authors declare no competing interests.\\

\bmhead{Data availability}

Source data are provided with this paper. Any additional data are available from the corresponding author upon request.\\

\bmhead{Code availability}

All codes to analyze the data in this study are available from the corresponding author upon request.\\

\begin{appendices}

\section*{Extended Data}
\setcounter{figure}{0}
\renewcommand\figurename{Extended Data Fig.}

\begin{figure}[hb]
\centering
\includegraphics[width=0.6\textwidth]{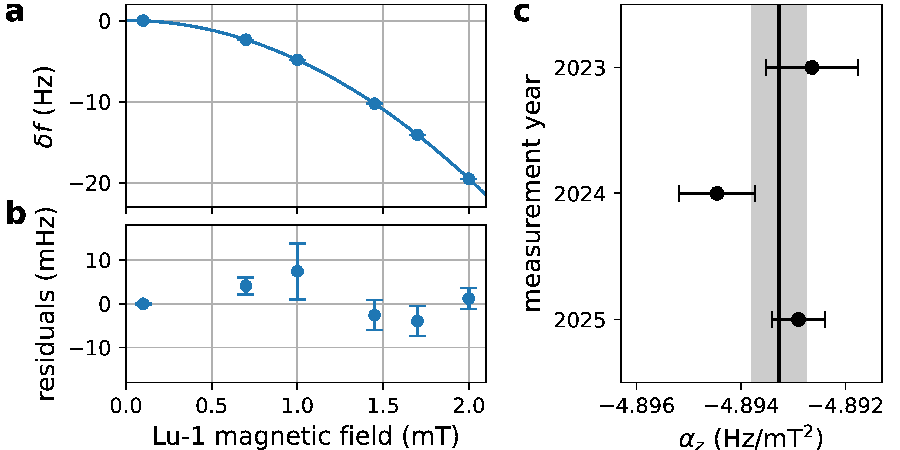}
\caption{Measurement of quadratic Zeeman coefficient. {\bf a,} measured frequency difference with Lu-1 operated over a range of field amplitudes and Lu-2 at 0.1 mT. The coefficient $\alpha_z$ is determined from a quadratic fit (solid line).  {\bf b,} Residuals with respect to the quadratic fit, which have a reduced $\chi^2_\nu=1.6$ for 5 degrees of freedom. {\bf c,} Three measurements of $\alpha_z$ over recent years.  The 2023 measurement was reported in~\cite{zhiqiang2023}, the 2024 measurement is reported here, and the 2025 measurement is from the data shown in {a,b}.  The vertical line and shaded area represent the combined value of $\alpha_z = -4.893\,27(51)\,\mathrm{Hz}/\mathrm{mT}^{2}$. }
\label{fig:qz}
\end{figure}

\begin{figure}[h]
\includegraphics[width=0.6\textwidth]{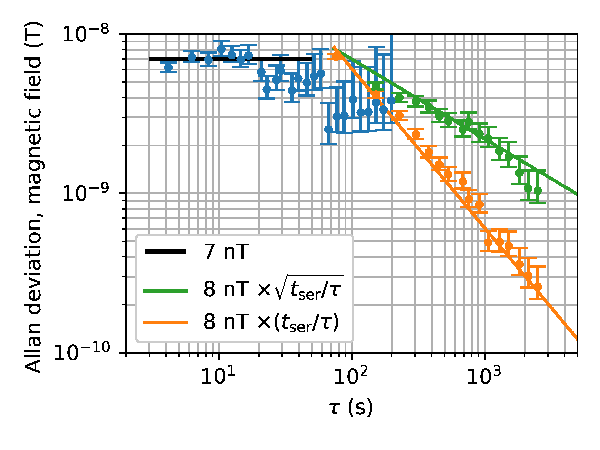}
\caption{Magnetic field instability. (blue) measurement of shorter time scale field instability without magnetic field compensation, (orange) instability inferred with magnetic field compensation from in-loop $\ket{^3D_1,6,\pm1}$ Zeeman splitting measurements, and (green) independent measurement of magnetic field instability with magnetic field compensation but inferred from out-of-loop $\ket{^3D_1,8,\pm1}$ Zeeman splitting measurements.}
\label{fig:mag}
\end{figure}

\begin{figure}[h]
\includegraphics[width=0.6\textwidth]{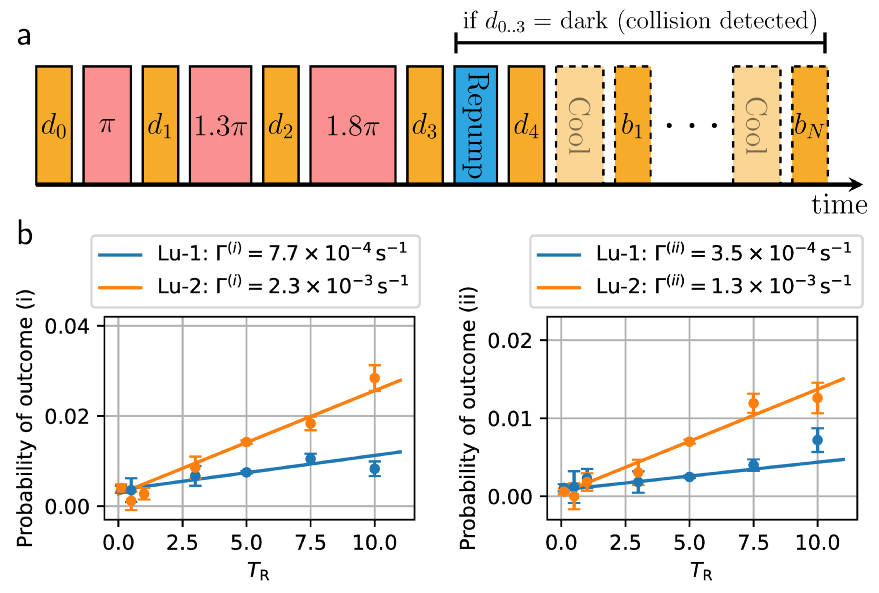}
\caption{{\bf a} Detection sequence: (orange) $d_i$ are standard Bayesian state detection and $b_i$ high threshold fixed time detection, (red) 848 nm clock pulses of variable area for shelving, (blue) repump (350 nm, 622 nm, and 895 nm) pulse, and (light orange) laser cooling. {\bf b} Probability of occurrence for outcome (i), first detected bright at $d_4$, and outcome (ii), first detected bright on $b_i$, as a function of Ramsey time for Lu-1 (blue points) and Lu-2 (orange points). Solid lines are linear fits to determine rates.}
\label{fig:collision}
\end{figure}

\begin{figure}[h]
\includegraphics[width=0.48\textwidth]{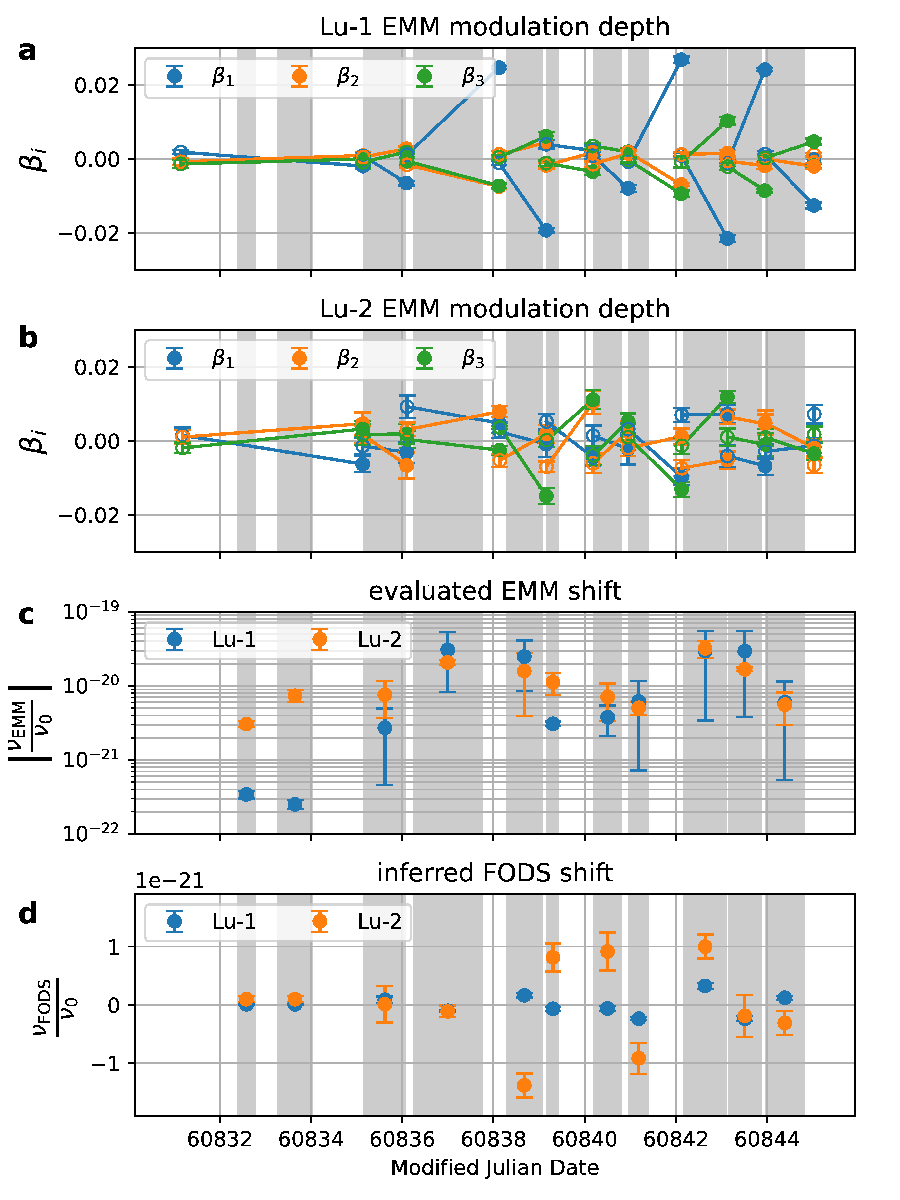}
\caption{{\bf a-b,} EMM modulation depths $\beta_{m,i}$ measured in three orthogonal directions. Solid circles are measurements before compensation, and open circles are measured immediately after compensation. Solid lines are an interpolation over measurement durations. {\bf c} Total EMM shift evaluated for each comparison measurement interval. {\bf d} First-order Doppler shift (FODS) evaluated for the ion displacements along the probe direction inferred from $\beta_{m,3}$ measurements.}
\label{fig:emm}
\end{figure}

\begin{figure}[h]
\includegraphics[width=0.6\textwidth]{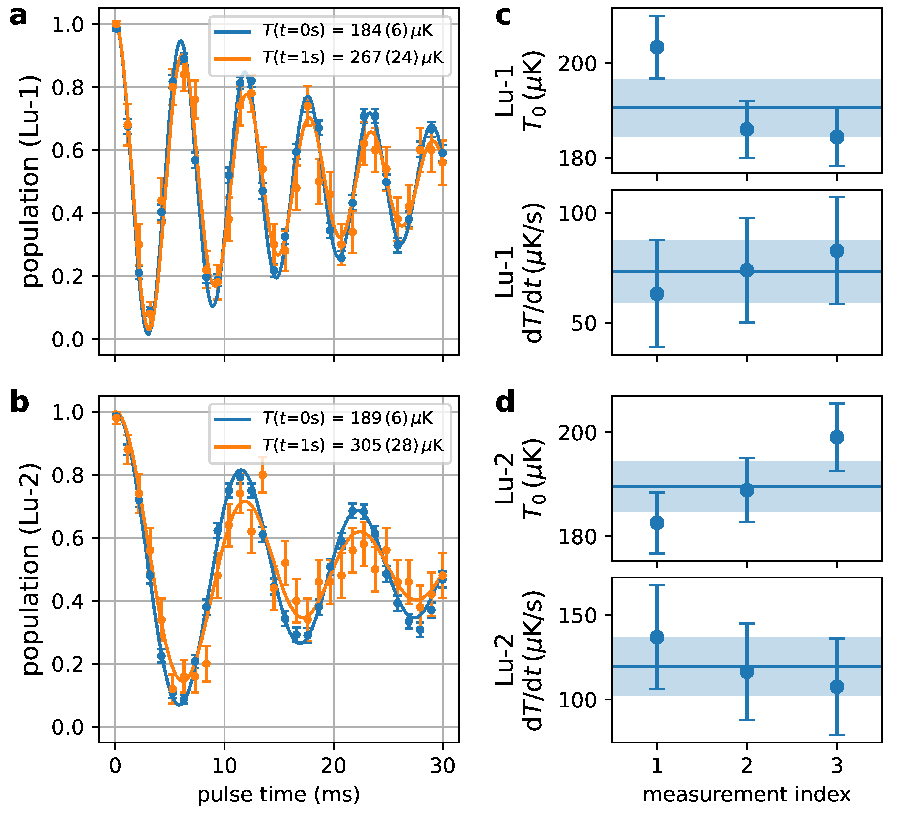}
\caption{{\bf a-b,} A typical measurement of the axial temperature by thermal dephasing after Doppler cooling $(t=0)$ and after a 1 second delay. {\bf c-d} Initial axial temperature and heating rates from three measurements over the course of the campaign for Lu-1 and Lu-2.}
\label{fig:axtherm}
\end{figure}

 \begin{figure}[h]
\includegraphics[width=0.6\textwidth]{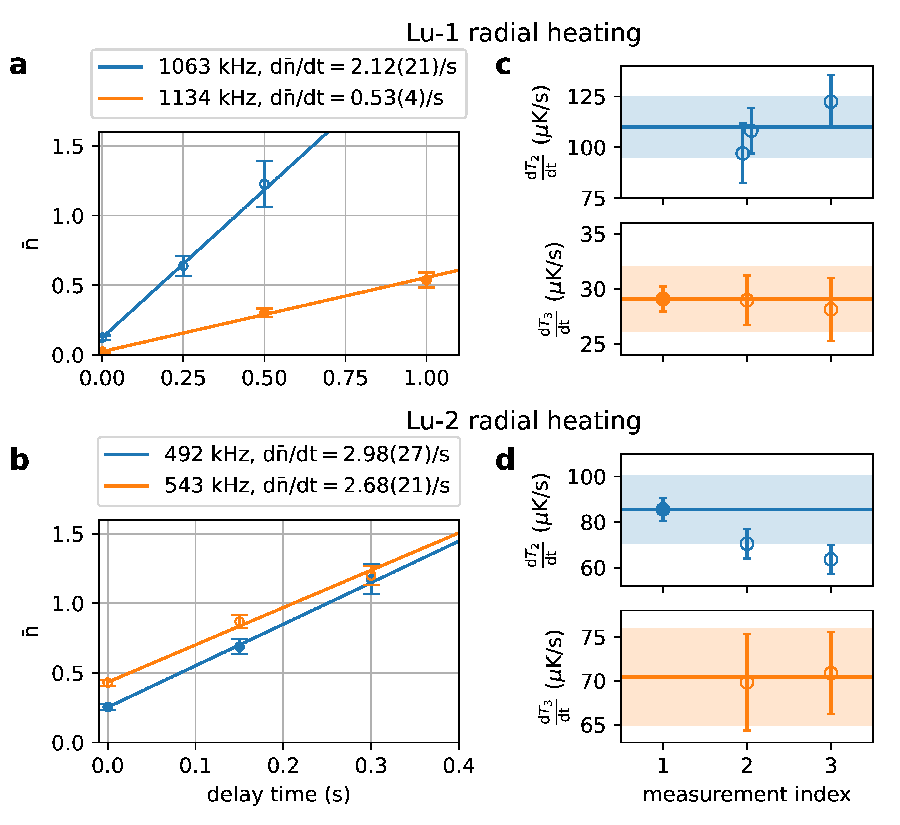}
\caption{Radial heating rates for Lu-1 ({\bf a,c}) and Lu-2 ({\bf b,d}) measured by sideband ratio thermometry after ground state cooling. {\bf a,b} are representative heating rate measurements for both radial modes in each trap. {\bf c,d} Heating rates obtained from all repeat measurements. Solid circles (index 1) were taken at the time of the measurement campaign; and open circles one year later with the same trap operating parameters, with measurements at indices 2 and 3 separated by approximately one week. Solid lines and shaded areas represent the radial heating rates and uncertainties used for SODS shift evaluation. For a radial mode measured during the campaign, the campaign value is used; for the remaining modes, the weighted mean of the later measurements is used. In all cases the uncertainty is taken as the largest among the individual measurements of that mode.  }
\label{fig:radtherm}
\end{figure}

 \begin{figure}[h]
\includegraphics[width=0.6\textwidth]{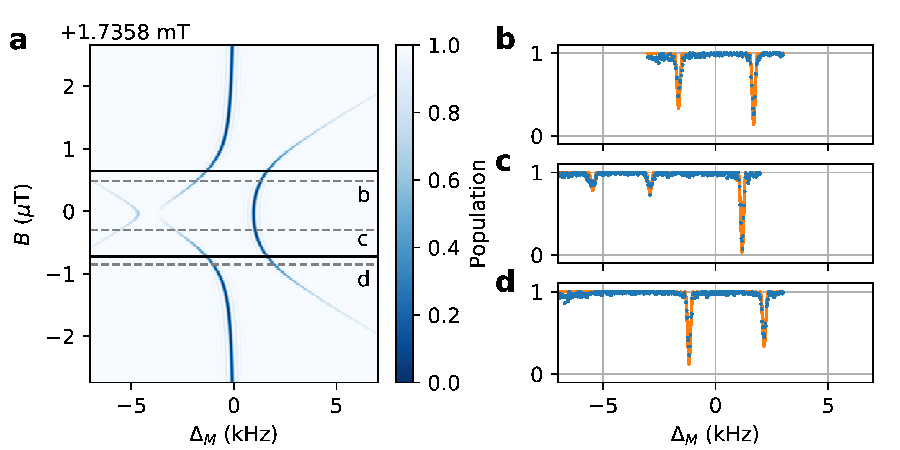}
\caption{{\bf a} Simulated Autler-Townes splitting for $\Omega_M = 2\pi\times 100\,$Hz and $\Omega_B = 2\pi\times 3\,$kHz in Lu-1. Solid black lines denote the Autler-Townes resonance conditions $\Delta_{\pm1} = 0$. Dashed lines indicate the magnetic fields corresponding to experiment scans {\bf b-d}. {\bf b-d} Simulation results (orange lines) of Hamiltonian \eref{eqn:AT_H}  which are fit to experimental data (blue points) with $\Omega_B$ as a free parameter.}
\label{fig:autler}
\end{figure}

 \begin{figure}[h]
\includegraphics[width=0.48\textwidth]{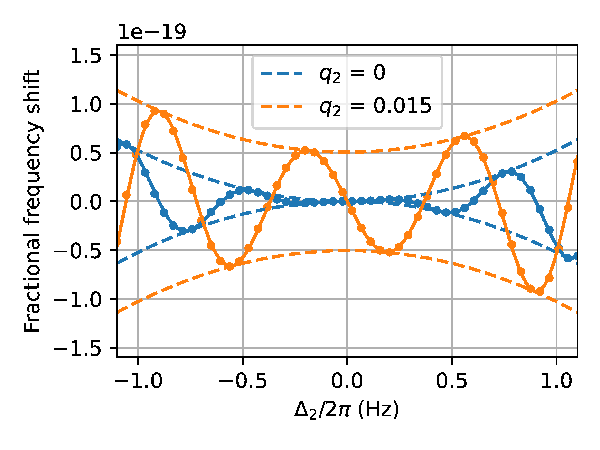}
\caption{Fractional frequency shift for HA-HR due to microwave coupling errors.  Evaluated with $T=5\,$s, $\tau_L=8\,$ms, and $\tau_1=\tau_2=12\,$ms as function of microwave detuning $\Delta_2$ for both no coupling error ($q_2 = 0$, blue) and 1.5\% error ($q_2 = 0.015$, orange). Points are exact numerical simulation, solid lines are evaluated by the analytic approximation given in the Supplementary Material of~\cite{zhiqiang2023} and dashed lines are the bounding envelope for the uncertainty given by \eref{eq:eerr}. }
\label{fig:uwcouplingerror}
\end{figure}

 \begin{figure}[h]
\includegraphics[width=0.48\textwidth]{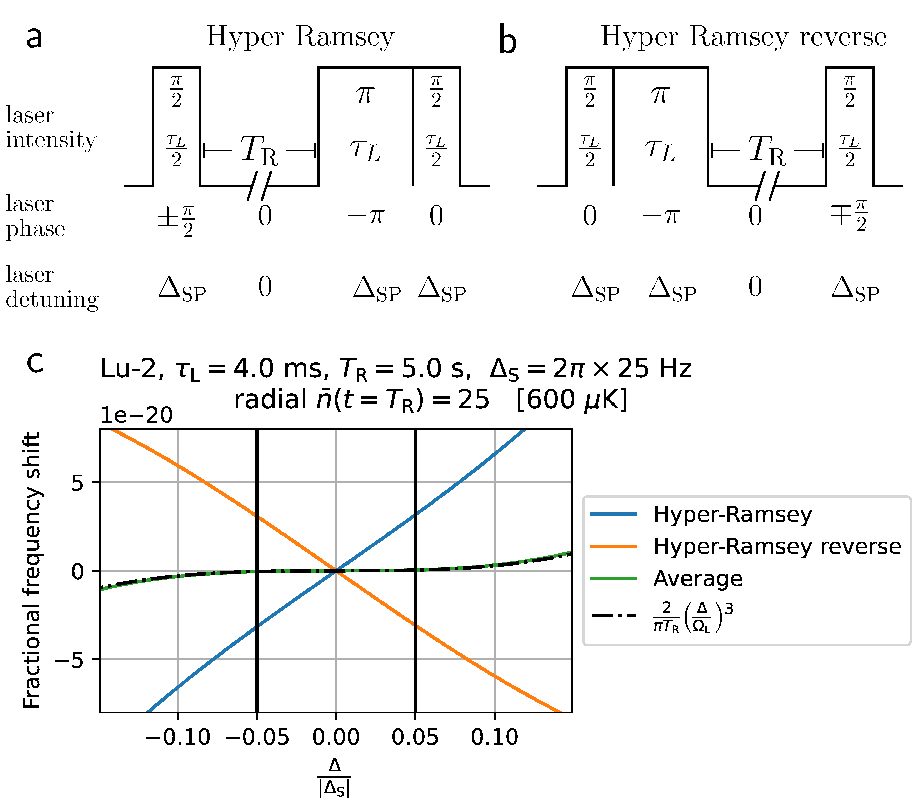}
\caption{{\bf a-b} Hyper-Ramsey and Hyper-Ramsey reverse pulse sequences. {\bf c} Simulation of Hyper-Ramsey suppression when including thermal effects for both sequences for the typical operating parameters of Lu-2. By averaging the two sequences, one recovers the cubic sensitivity of Hyper-Ramsey independent of the final thermal state.}
\label{fig:hyperramsey}
\end{figure}

 \begin{figure}[h]
\includegraphics[width=0.6\textwidth]{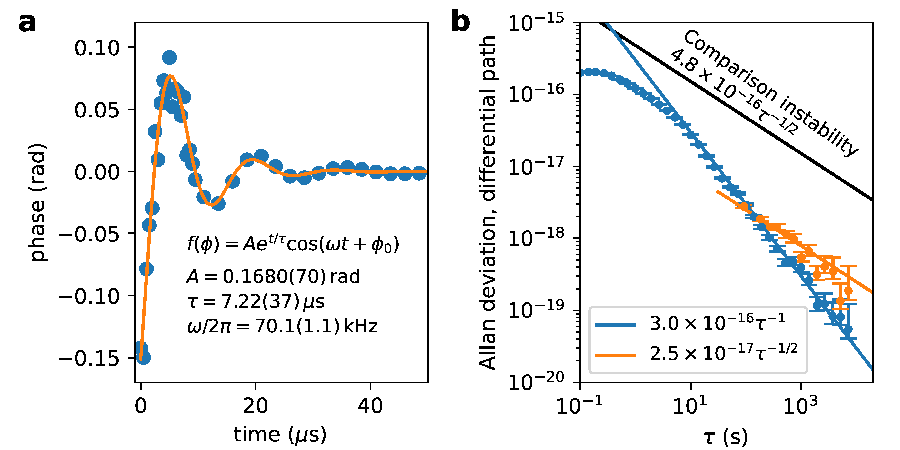}
\caption{{\bf a} Measured differential phase chirp when switching the AOMs for optical pulses. The phase chirp is modeled as a damped harmonic oscillation with fit parameters given in the plot.  {\bf b} Blue points are instability from differential optical phase fluctuations measured out-of-loop with an additional 2 m of unstabilized path while the path length stabilization is engaged. Orange is the estimated contribution during comparison measurements due to sampling at 87\% duty cycle.}
\label{fig:fods}
\end{figure}

\end{appendices}

\clearpage

\includepdf[pages=-, scale=0.9, pagecommand={}]{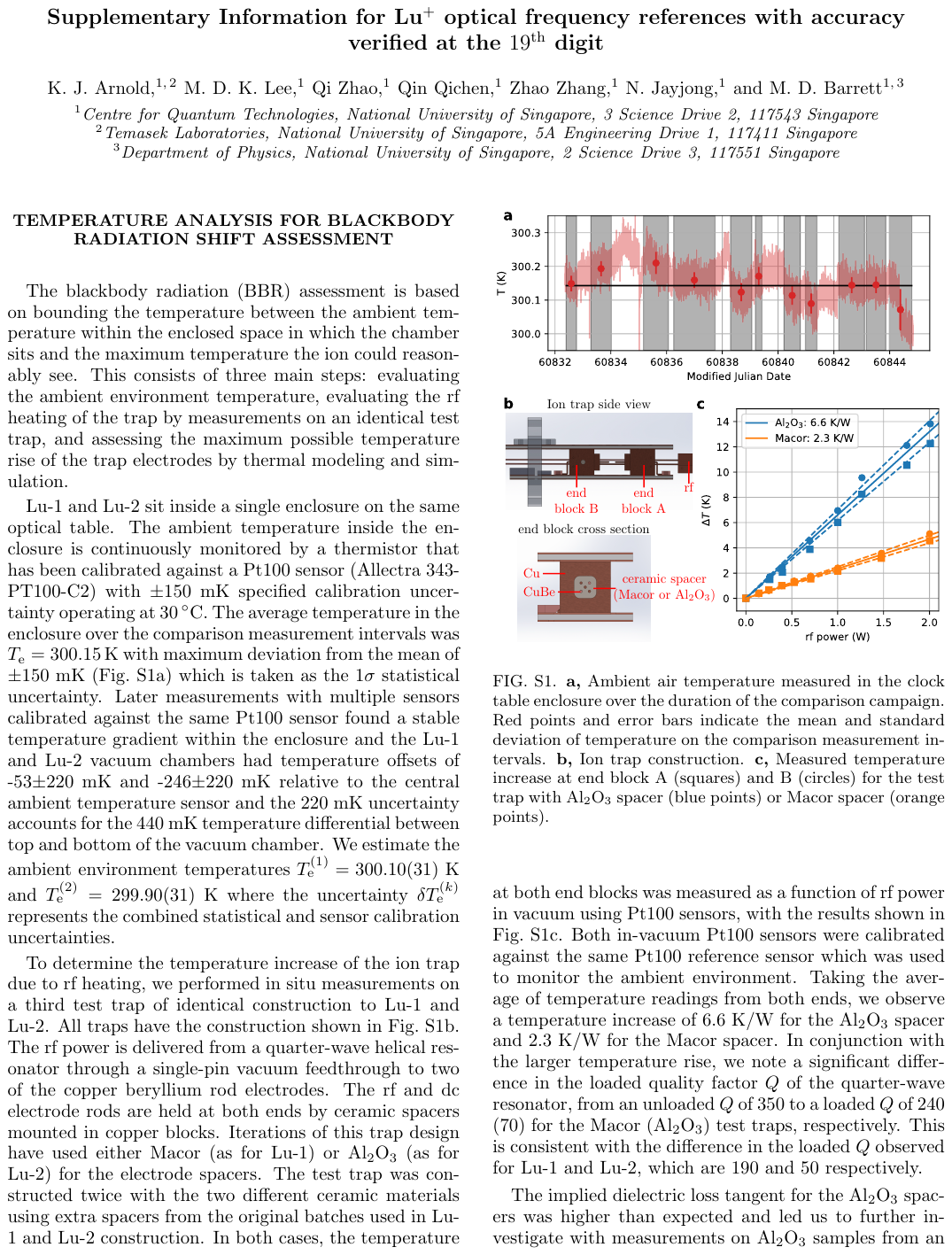}

\end{document}